\documentclass{aa}  
\usepackage{graphicx}
\usepackage{txfonts}
\usepackage{natbib}

\begin{document}

   \title{One dimensional prominence threads}

   \subtitle{I. Equilibrium models}

   \author{J. Terradas
          \inst{1}, M. Luna\inst{1}, R. Soler\inst{1}, R. Oliver\inst{1}, M. Carbonell\inst{2}, J. L. Ballester\inst{1}, 
          }

   \institute{$^1$Departament de F\'\i sica, Universitat de les Illes Balears (UIB),
E-07122, Spain \\   Institute of Applied Computing \& Community Code (IAC$^3$),
UIB, Spain\\ \email{jaume.terradas@uib.es}\\
$^2$Departament de Ci\`encies Matem\`atiques i Inform\`atica, Universitat de les Illes Balears (UIB),
E-07122, Spain\\   Institute of Applied Computing \& Community Code (IAC$^3$),
UIB, Spain
}

   \date{}

 
  \abstract
   {Threads are the building blocks of solar prominences and very often 
   show longitudinal oscillatory motions that are strongly attenuated with time.
   The damping mechanism responsible for the reported oscillations is not fully
   understood yet.}
   {To understand the oscillations and damping of prominence threads it is mandatory to
   investigate first the nature of the equilibrium solutions that arise under
   static conditions and under the presence of radiative losses, thermal conduction and background heating. This provides the basis to calculate the eigenmodes of the
   thread models.}
   {The nonlinear ordinary differential equations for hydrostatic and thermal
   equilibrium under the presence of gravity 
   are solved using
   standard numerical techniques and simple analytical expressions are
   derived under certain approximations. The solutions to the equations represent a prominence thread, i.e., a
   dense and cold plasma region of a certain length that 
   connects with the corona through a prominence corona transition region (PCTR).
   The solutions can also match with a chromospheric-like layer if a spatially dependent heating function
   localised around the footpoints is 
   considered.}
   {We have obtained static solutions representing prominence threads and have
   investigated in detail the dependence of these solutions on the different
   parameters of the model. Among other results,  we have shown that multiple condensations along a magnetic field line are possible, and that the effect
   of partial ionisation in the model can
   significantly modify the thermal balance in the thread and therefore their
   length. This last parameter is also shown to be comparable to that reported in the
   observations when the radiative losses are reduced for typical thread
   temperatures.}
   {}

   \keywords{Magnetohydrodynamics (MHD) --- waves --- Sun: magnetic fields
               }

   \maketitle
%

\section{Introduction}\label{introduction}

A recent survey on longitudinal oscillations in solar filaments carried out by
\citet{lunaetal2018} has provided interesting results about the temporal
attenuation of the oscillatory motions. A measure of the attenuation is the
ratio between the damping time, assuming an exponential decay, and the period of
the oscillation. The mean value of this parameter over 106 small amplitude
events (with velocities below 10 km $\rm s^{-1}$) is 1.75, while for large
amplitude oscillations (above 10 km $\rm s^{-1}$) the 96 events give a mean
value of 1.25. This means that typically the oscillations do not last more than
2 periods. The question that arises is the mechanism that produces such strong
damping, and several approaches can be used to investigate this problem. 

The first approach is to consider that the system is near equilibrium satisfying
the energetic balance between radiation losses, thermal conduction and heating.
The model may include the effect of the magnetic field and also the
gravitational force. Once the equilibrium is calculated, and this is the main
motivation of the present work (Paper I), the problem of linear and nonlinear waves on
these equilibrium configurations can be then addressed (Paper II). Only a few
works have focused on the determination of a static equilibrium under thermal
balance:  \citet{degendeinzer1993} modelled a quiescent prominence assuming
balance between heating and radiative losses but ignoring heat conduction.
These authors found reasonable values for prominence temperatures and densities
but significantly shorter threads were obtained \citep[a similar result was
obtained by][]{milne1979}. The connection with the
chromosphere was not included in the model. Later, \citet{dahlburgetal1998}
demonstrated the role of a dipped geometry to support a prominence condensation
against gravity and included a localised footpoint heating to match their
solution with a chromospheric layer. Regarding the oscillations 
\citet{schmittdegen1995} and \citet{rempeletal1999} performed a stability
analysis of a flux tube model based on the work of \citet{degendeinzer1993}
under line-tying conditions. However, in these studies the perturbations were
assumed to be adiabatic and therefore no attenuation was reported, although some
hints of instability were found. The inclusion of nonadiabatic effects under
different boundary conditions will be addressed in Paper II. In addition, the effect
of partial ionisation, effective for plasma temperatures below 10,000 K, has not
been taken into account in the above mentioned studies.

The second approach to investigate thread or prominence oscillations is based on
the analysis of a dynamical system that undergoes thermal nonequilibrium.
Oscillations are naturally produced in the system and their investigation
provides details about the damping processes. Along this line, \citet{lunakarpen12} investigated the oscillations of multiple threads formed in long, dipped flux tubes through the thermal nonequilibrium process previously simulated \citep{lunaetal12a}. These
authors found that the oscillation properties predicted by their simulations are
in agreement with the observed behaviour, and that the main restoring force is
the projected gravity along the tube where the threads oscillate.
\citet{zhangetal2012,zhangetal2013} modelled impulsive heating at one leg of the
loop and an impulsive momentum deposition, which cause the prominence to
oscillate. The oscillation damps with time under the presence of non-adiabatic
processes and these authors concluded that radiative cooling is the dominant
factor leading to damping. 

In the present paper we follow the first approach. We construct improved static
prominence/thread models by considering different forms of the radiative losses under
the presence of gravity. The basic conditions to have a temperature minimum at the
thread centre are derived, and the possibility of having several condensations along the
field line is discussed. The connection of the thread solution with the prominence corona transition region (PCTR) and
eventually with a chromospheric layer near the footpoints is also investigated.
Partial ionisation is included in the model and its effect on the obtained solutions
is studied. The work presented here is the basis of the analysis of the
eigenmodes described in the forthcoming Paper II. 

\section{Basic model}

\begin{figure}[!hh] \centering \includegraphics[width=9.25cm]{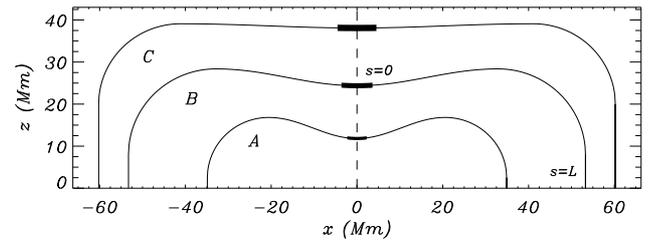} 
\caption{\small Sketch of the assumed magnetic field geometry. The different
parameters allow to change the shape and curvature of the field lines where the
thread is allocated \citep[based on][]{dahlburgetal1998}. The configurations are
denoted by $A$, $B$ and $C$. The three configurations have been represented with
different total lengths for visualisation purposes only.  The  parameters of
each curve following the definition in \citet{dahlburgetal1998} are A (bottom
curve):  $s_1 = 0.05 L$, $s_2 = 0.5 L$, $d= 0.1 L$, B (middle curve): $s_1 = 0.1
L$, $s_2 = 0.5 L$, $d= 0.05 L$, and C (top curve): $s_1 = 0.2 L$, $s_2 = 0.5 L$,
$d= 0.01L$. $L$ is half the length of the magnetic field
line.}\label{fieldlines} \end{figure}

We assume that the magnetic field dominates the plasma and reduce
the problem to a one-dimensional thread in which the magnetic field determines
the geometry but is not modified by the presence of the thread. The plasma
quantities change along the thread and therefore we concentrate on the forces and equilibrium
along the field only.  This assumption is valid for prominences embedded in sufficiently strong magnetic fields \citep[e.g.][]{fiedlerhood92,hilliervan13,jenkinsetal2019}. For a model with equilibrium normal to the magnetic field
see for example, \citet{ballesterpriest1989}. For simplicity here we concentrate
on a 1D problem and the tube cross-section is assumed to be constant. We adopt a symmetric geometry
about the midpoint with a dipped region in the central part as in
\citet{antiochosklimchuk1991}, \citet{dahlburgetal1998} and
\citet{zhouetal2014}. The definition of the field geometry is given by
Eq.~(6) of \citet{dahlburgetal1998}. We denote by $g(s)$ the gravity
acceleration parallel to the field line and $s$ the distance along this line
(starting at the midpoint of the structure, i.e., the centre of the thread). 
Three different examples of field geometry investigated in the present work are represented in
Fig.~\ref{fieldlines}.

\section{Fully ionised plasma}\label{equil}

For a static situation two
equations for the force balance and thermal equilibrium must be satisfied
simultaneously. Gas pressure under static equilibrium must satisfy that 
\begin{eqnarray}\label{hydrop}
\frac{dp}{ds}(s)=\rho(s) g(s).
\end{eqnarray}
When gravity is neglected gas pressure remains constant
along $s$. This equation is completed with the ideal gas law 
\begin{eqnarray}\label{gaslaw} p(s)=\frac{1}{\tilde{\mu}} \frac {k_{\rm B}}{m_{\rm p}}\rho(s) T(s), \end{eqnarray}
where $k_{\rm B}/m_{\rm p}$ is the ideal gas constant and $\tilde{\mu}$ is the mean atomic weight.
For a fully ionised hydrogen plasma $\tilde{\mu}=1/2$. The modifications in the previous equation because of partial ionisation will be 
introduced later in Sect.~\ref{partialion}.

Using the ideal gas law we can eliminate the variable $p(s)$ to obtain,
\begin{eqnarray}\label{hydro}
\frac{d\rho}{ds}(s) T(s)+ \rho(s) \frac{dT}{ds}(s)=\tilde{\mu} \frac{m_{\rm p}}{k_{\rm B}}\rho(s) g(s).
\end{eqnarray}
For thermal equilibrium between conduction, radiative losses and heating the next nonlinear ordinary differential equation has to be satisfied,
\begin{eqnarray}\label{thermal}
\frac{d}{ds} \left(\kappa_\parallel(s)
\frac{dT}{ds}(s)\right)-\rho(s)^2 \Lambda(T(s))+E_0=0,
\end{eqnarray}

\noindent where $\kappa_\parallel(s)= \kappa_0\, T(s)^{5/2}$, with $\kappa_0$ a 
constant coefficient.  $\Lambda(T)$ is the radiative loss function and
$E_0$ is the background heating. It is instructive to write the previous energy equation in
integral form using the Gauss theorem in one dimension,
\begin{eqnarray}\label{flux} \left[\kappa_\parallel(s) \frac{dT}{ds}(s)\right]_{s=\pm
L}+\int_{-L}^{L}\left(-\rho(s)^2 \Lambda(T(s))+E_0\right)ds=0, \end{eqnarray} where
the first term of this equation represents the (minus) heat flux through the footpoints of the
field lines located at $\pm L$ and the second term is the spatially integrated
contribution of radiation and heating. If $E_0$ is zero the heat flux through
the boundaries because of thermal conduction, is equal to the total energy lost by
radiation in the whole domain. 

\begin{figure}[!hh] \center{\includegraphics[width=9.5cm]{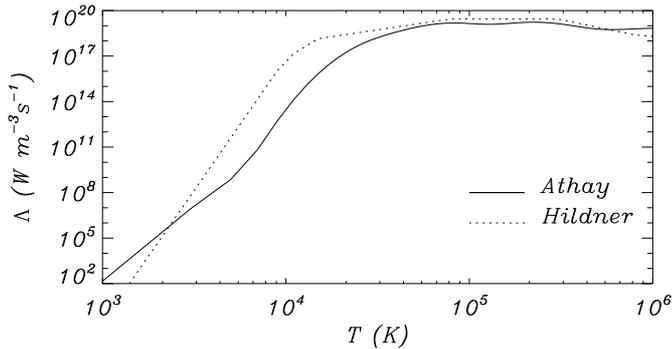}} 
\caption{\small Radiative loss functions used in the present work,  \citet{athay1986} (continuous line) and 
\citet{hildner1974} (dotted line). Athay's function has reduced losses in comparison with Hildner for typical temperatures in the range 4,000-12,000 K.}\label{equilrad} \end{figure}

For optically thin radiative losses the function $\Lambda(T)$ has several
parametrisations in the literature that are essentially valid for the corona
surrounding prominences. Nevertheless, threads are optically thick and therefore
radiative losses are expected to be greatly reduced. This effect can be taken
into account by artificially  decreasing the values in the optically thin
radiative losses for low temperatures ($T<10^4 K$), see for example
\citet{milne1979}, \citet{schmittdegen1995}. Here we consider two main radiative
loss functions, described in  \citet{athay1986} and \citet{hildner1974} and represented in Fig.~\ref{equilrad} for comparison purposes. Athay's
radiation function has the advantage of being an analytical function with
continuous derivatives and attains quite low values under prominence/thread
conditions. As we show in the next sections this has important consequences
regarding  the lengths  of the calculated threads. We have explored other
radiative losses such as Klimchuk-Raymond's fits \citep{klimchukcarg2001} and the parametrisation
computed from CHIANTI atomic database \citep{dereetal1997,landietal2012} but the results are in general quite
similar to those of Hildner's function.

We derive the conditions to obtain a dense and cold thread at the centre of
the dip surrounded by a hot plasma. We denote the temperature and density at this point as $T_0=T(0)$ and $\rho_0=\rho(0)$. By symmetry around $s=0$ we
impose that $dT/ds=0$, meaning that Eq.~(\ref{thermal}) at $s=0$ reduces
to 
\begin{eqnarray}\label{thermals0}
\frac{d^2 T}{ds^ 2}(0)= 
\frac{\rho_0^2\, \Lambda(T_0)-E_0}{\kappa_\parallel(0)}.
\end{eqnarray}

\noindent Since we are seeking for solutions that represent a cold thread
connecting with the hot corona, the temperature at $s=0$ (where there is an
extrema because we have imposed that $dT/ds=0$) must have a minimum, i.e., $d^2
T/ ds^2>0$. According to Eq.~(\ref{thermals0}) this condition is satisfied
only if $E_0<\rho_0^2\,\Lambda(T_0)$. Conditions for the existence of
prominence solutions were already studied in some detail in the early work of
\citet{milne1979}. Note that even in the unlikely situation with no background heating in the solar atmosphere, i.e.,
$E_0=0$, it is still possible to obtain physically meaningful solutions.
However, for $E_0>\rho_0^2\,\Lambda(T_0)$ no solutions representative of threads
(i.e., cold material surrounded by coronal plasma) are found. Instead, coronal
loop solutions of the type studied by, for example, \citet{klimchuketal2010} or
\citet{mikicetal2013} are obtained. In fact, the stationary solutions calculated
by these last authors by solving the time dependent problem have been used as a
check of our numerical method based on standard routines. We have obtained static equilibrium solutions that match well with the stationary solutions of \citet{mikicetal2013}.

\begin{figure}[!hh] \centering \includegraphics[width=8cm]{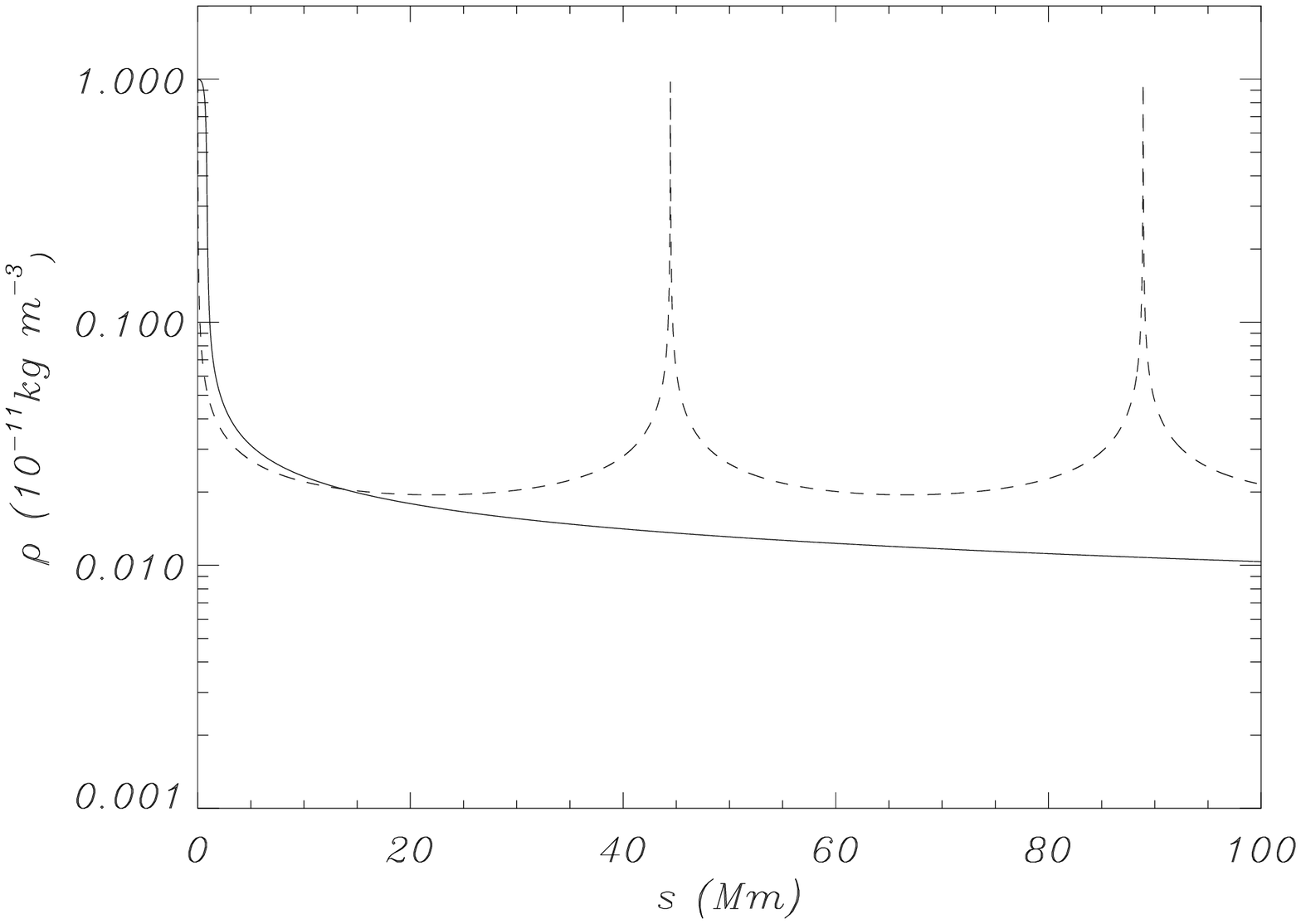}
\includegraphics[width=8cm]{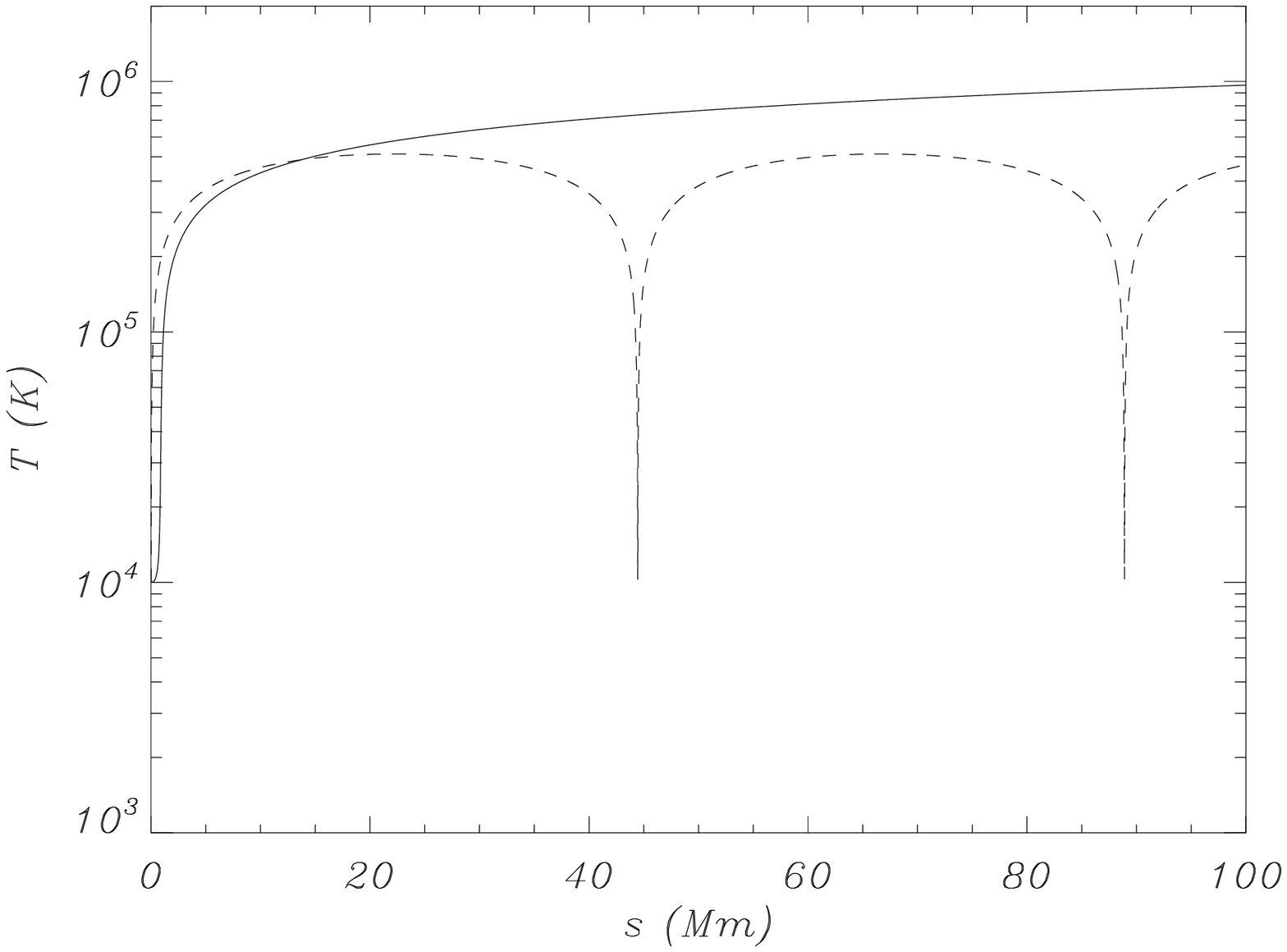} \caption{\small Hydrostatic and thermal
equilibrium along the field line with zero gravity. The continuous line corresponds
to Athays's radiation function, the dashed line represents the results for
Hildner's function, while the background heatings are $E_0=2.75\times 10^{-9}$ W
m$^{-3}$ and  $E_0=1.75\times 10^ {-6}$ W m$^{-3}$. In this
particular example $T_0=10^4\,\rm K$,  $\rho_0=
10^{-11}\,\rm kg \, m^{-3}$ and the total length of the field line is $2 L=200\,\rm Mm$. Model $A$ is used in this plot.}\label{comparg0} \end{figure}

Typical numbers for background heating found in the literature are in the range $10^{-4}-10^{-5}$ W m$^{-3}$ \citep[e.g.,][]{karpenetal2001,mikicetal2013} however, for the specific radiative losses considered in this work, these values are in general too high to satisfy the condition  $E_0<\rho_0^2\,\Lambda(T_0)$ necessary to have a typical thread/prominence with density $\rho_0= 10^{-11}$ kg m$^{-3}$ and temperature $T_0=10,000$ K. For this reason, we have decided not to focus on a single value of the background heating but changing the value of $E_0$ in the range between 0 and slightly below  $\rho_0^2\,\Lambda(T_0)$. It is clear that the value of the radiative loss function at $T_0$ is a relevant number that determines the range of allowed values for $E_0$. If the radiative losses are low for temperatures around  $10,000$ K the background heating needs to be decreased accordingly. For this reason we consider background heatings as low as $E_0=2.75\times 10^{-9}$ W
m$^{-3}$ and  $E_0=1.25\times 10^ {-6}$ W m$^{-3}$ for Athay and Hildner's radiation functions but in some cases the values are slightly above these values (e.g., Fig.~\ref{comparg0}).

The system of Eqs.~(\ref{hydro}) and
(\ref{thermal}) are solved numerically for $T$ and $\rho$ under given boundary
conditions using standard numerical techniques based on a variable-order, variable-step Adams method. At the centre of the thread ($s=0$)  the temperature ($T_0$) and
density ($\rho_0$) values are selected and $dT/ds=0$ is imposed.  The  two
coupled ordinary differential equations are numerically integrated from $s=0$ to
$s=L$ using numerical methods. When necessary, adaptive mesh methods
are used, this is especially important at the PCTR. We do not need, in general, to implement a shooting method since the
three boundary conditions are imposed at $s=0$. The shooting method is only
required when density at $s=0$ has to be determined to match the
temperature at the corona ($s=L$), see end of Sect.\ref{paramsurv}. In this last case we use a Runge–Kutta–Merson method and a Newton iteration in the shooting and matching technique.

\subsection{Constant gas pressure}

For a better comprehension of the results we start with the case with zero
gravity. Gas pressure is constant along the field under such conditions, and the
solution of the equations is represented in Fig.~\ref{comparg0} for the two
radiation functions. The solutions show a rather different behaviour although
the reference temperatures and densities are the same at $s=0$. For Athay's
radiation function a single thread around $s=0$ is found and matches smoothly through a 
PCTR a plasma that is close to coronal
conditions (densities of the order of $10^{-13}\,\rm kg\, m^{-3}$, and
temperatures around $10^6\,\rm K$). On the contrary, in the case of Hildner's
function, several (up to three) cold and dense regions are found in the
configuration. The densities and temperatures of these thread-like solutions are chosen to be 
exactly the same at the centre of the thread ($s=0$). The system displays periodic cold and dense threads, and this
behaviour also applies to Athay's radiation function but the corresponding
spatial periodicity is much longer than the length of the system. 

The distance between successive threads depends on the reference values for
temperature and density and also on the value of the heating constant. It is
clear that for constant gas pressure the system has a characteristic spatial scale and it
is worth investigating the origin of this length. It turns out that this scale
is closely related to the so-called Field's  wavenumber after \citet{field1965}. This
author found that the thermal mode in a uniform plasma with constant temperature of the
order of MK is always unstable if thermal conduction is absent. But under the
presence of thermal conduction there is a critical length scale, that we
denote hereafter by
$L_{\rm C}$,
for the stabilisation of the thermal mode. In the model used by \citet{field1965}, with
uniform temperature and density, the equilibrium satisfies that
\begin{eqnarray}\label{equileq}
\rho_0^2\,\Lambda(T_0)=E_0,
\end{eqnarray}
and for a given constant pressure value and using the gas law, density and temperature are obtained from the
previous equation, which must be solved numerically in general. Since the
equilibrium is isothermal there is no conduction term in Eq.~(\ref{equileq}) but it is present in the perturbations.  

An analysis of the perturbations in this configuration to understand the features of the thermal mode has been done in the past by, for example, \citet{field1965,vandergoossens1991,soleretal11a,soleretal2012}. The obtained dispersion relation shows the existence of a critical length and only those perturbations with wavelenghts below $L_{\rm C}$ are stable, the expression for $L_{\rm C}$ \citep[see Eq.~(26a) in][]{field1965} is
\begin{eqnarray}\label{lfield}
L_{\rm C}&=&2\pi\sqrt{
\frac{{\kappa_\parallel}}{{\frac{\rho}{T}}\left(\frac{\partial \rho^2
\Lambda}{\partial \rho}\right)_T-\left(\frac{\partial \rho^2 \Lambda}{\partial
T}\right)_\rho}}\nonumber\\&=&2\pi\sqrt{
\frac{{\kappa_0\,T_0^{5/2}}}{2\frac{\rho_0^2}{T_0}
\Lambda(T_0)-\rho_0^2\,\Lambda^{\prime}(T_0)}}. \end{eqnarray}
Derivatives involving density and radiative losses at constant temperature
and constant density are present in the denominator of the previous expression and are evaluated in the second part of the equation where $\Lambda^{\prime}(T_0)$ is the temperature derivative of the radiative losses evaluated at $T_0$. It is worth to mention that in Eq.~(\ref{lfield}) the explicit dependence on $E_0$ is absent, the reason being that in our model the heating only affects the equilibrium but not the perturbations. 
Note that \citet{begelmanmackee1990} define a Field's length which is not the same as the critical length  used here \citep[see also][]{koyamainut2004,sharmaetal2010}. \citet{begelmanmackee1990} provide a relationship between the critical and the Field's length in their Eq.~(4.16).

We apply Eq.~(\ref{lfield}) to the situation in
Fig.~\ref{comparg0} but keeping in mind that we are going to compare the calculated inhomogeneous equilibrium with a model that has constant density and temperature. In order to perform a reliable comparison we chose the same gas pressure value (which is constant along the tube) and the same background heating in the two models. Then  we calculate the corresponding density and temperature that satisfies
Eq.~(\ref{equileq}) for the homogeneous model. For example, for Hildner radiation function we find a temperature of $342536\,\rm K$ and a density of $2.9\times 10^{-13}\,\rm kg\, m^{-3}$ in the homogeneous model. If we compute the mean values for the corresponding inhomogeneous model (Fig.~\ref{comparg0}) we obtain a mean temperature of $418897\,\rm K$ and a mean density of $2.8\times 10^{-13}\,\rm kg\, m^{-3}$, these numbers are similar to those calculated for the homogeneous case. The obtained values of temperature and density for the homogeneous case are introduced in Eq.~(\ref{lfield}). We find that the corresponding scale length for Hildner radiation function is 36 Mm. This value agrees
reasonable well with the periodicity found in Fig.~\ref{comparg0} for
Hildner's function, which is around 43 Mm. Repeating the same procedure for Athay's radiation function (calculating again the temperature and density values) we find that in this case $L_{\rm C}= 22881$
Mm. This large value explains in Fig.~\ref{comparg0} the lack of periodicity in
a length of 100 Mm. Therefore, we conclude that the critical length provides
a reasonable estimation of the expected periodicity that makes the
system stable regarding the thermal instability. Furthermore, we have computed different numerical solutions by changing the value of $\kappa_0$ and have positively checked that
the obtained characteristic lengths are proportional to the square
root of the conduction
coefficient, as expected from Eq.~(\ref{lfield}). We have varied other parameters such as temperature and density and the results confirm that Eq.~(\ref{lfield}) is an adequate approximation. Note that the role of sound waves, and therefore, the effect of pressure variations is absent in the definition of the critical length while in the full numerical solutions of Fig.~\ref{comparg0} the effect of gas pressure is included.

Now we focus on another characteristic scale in the system but of local nature,
the length of the individual threads. As we have mentioned in
Section~\ref{introduction} the calculated lengths by
\citet{milne1979} and \citet{degendeinzer1993} are short in comparison with the
measured thread lengths. To investigate this question and since the dense plasma representing a thread is
located around the origin of our coordinate system we assume that the temperature around $s=0$ can be written as a second-order series expansion of the form
\begin{eqnarray}\label{tempexp}
T(s)=T_{0}\left(1+b_1\frac{s}{L}+b_2\frac{s^2}{L^2}\right),
\end{eqnarray}
where the dimensionless coefficients $b_1$ and $b_2$ need to be determined. Additional terms in Eq.~(\ref{tempexp}) are neglected since $s/L\ll 1$ and we perform a local analysis. The boundary condition $dT/ds=0$ at $s=0$ yields to $b_1=0$. For the plasma density we have that using the gas law for constant pressure and in the situation $s/L\ll 1$ we can write
\begin{eqnarray}\label{densexp}
\rho(s) \approx \rho_{0}\left(1-b_2\frac{s^2}{L^2}\right).
\end{eqnarray}
We substitute these expansions in $s$ for temperature and pressure in the full energy equation given by Eq.~(\ref{thermal}). Approximating the powers by the corresponding Taylor series and using again the fact that $s/L\ll 1$ we
find, after some algebra and to zeroth order in $s$ that
\begin{eqnarray}\label{bpar}
b_2=\frac{1}{2} \frac{L^2}{\kappa_0\, T^{7/2}_0} \left(\rho_0^2
\,\Lambda(T_0)-E_0\right).
\end{eqnarray}
This coefficient is in essence the ratio between the radiative minus the heating term over the conduction term evaluated at $s=0$ and assuming that the temperature changes on a spatial scale given by $L$. Equation~(\ref{bpar})
provides the value of the term in front of the parabolic dependence
on distance, $s$, in Eq.~(\ref{tempexp}) and it is used here as a proxy to obtain information about the thread length. The smaller the value of $b_2$ the flatter the parabolic curve and therefore the longer the length of the thread. The question that we have to address is how to define a spatial scale, $l_{\rm th}$, associated with a parabola. For this reason if we rewrite Eq.~(\ref{tempexp}) as 
\begin{eqnarray}\label{tempexp1}
T(s)=T_{0}\left(1+\frac{s^2}{l_{\rm th}^2}\right),
\end{eqnarray}
then
\begin{eqnarray}\label{lthread}
l_{\rm th} =\sqrt{\frac{L^2}{b_2}}=\sqrt{\frac{ 2\,\kappa_0\, T^{7/2}_0} {\rho_0^2\,\Lambda(T_0)-E_0}},
\end{eqnarray}
where we have used the expression for $b_2$ given in Eq.~(\ref{bpar}). The spatial scale $l_{\rm th}$, can be understood as an approximation for the thread length and provides valuable information about the dependence of this parameter on: central density and temperature,  radiative losses at this temperature, conduction and background heating. The radiative losses are dependent on the specific cooling table chosen as this
will define the behaviour and therefore shape of the PCTR. Since we are under the assumption $E_0<\rho_0^2\,\Lambda(T_0)$, the parameter 
$l_{\rm th}$ is always a real positive number. According to Eq.~(\ref{lthread}) 
an increase in the conduction will lead to longer threads while an increase in
the radiative losses will shorten them (if $E_0$ is constant). The dependence of $l_{\rm th}$ on the conduction coefficient, $\kappa_0$, is exactly the same as for $L_{\rm C}$ (see Eq.~(\ref{lfield})) although these two spatial scales have different physical meanings. We will return to Eq.~(\ref{lthread}) in the forthcoming sections.



\begin{figure}[!hh] \centering \includegraphics[width=8cm]{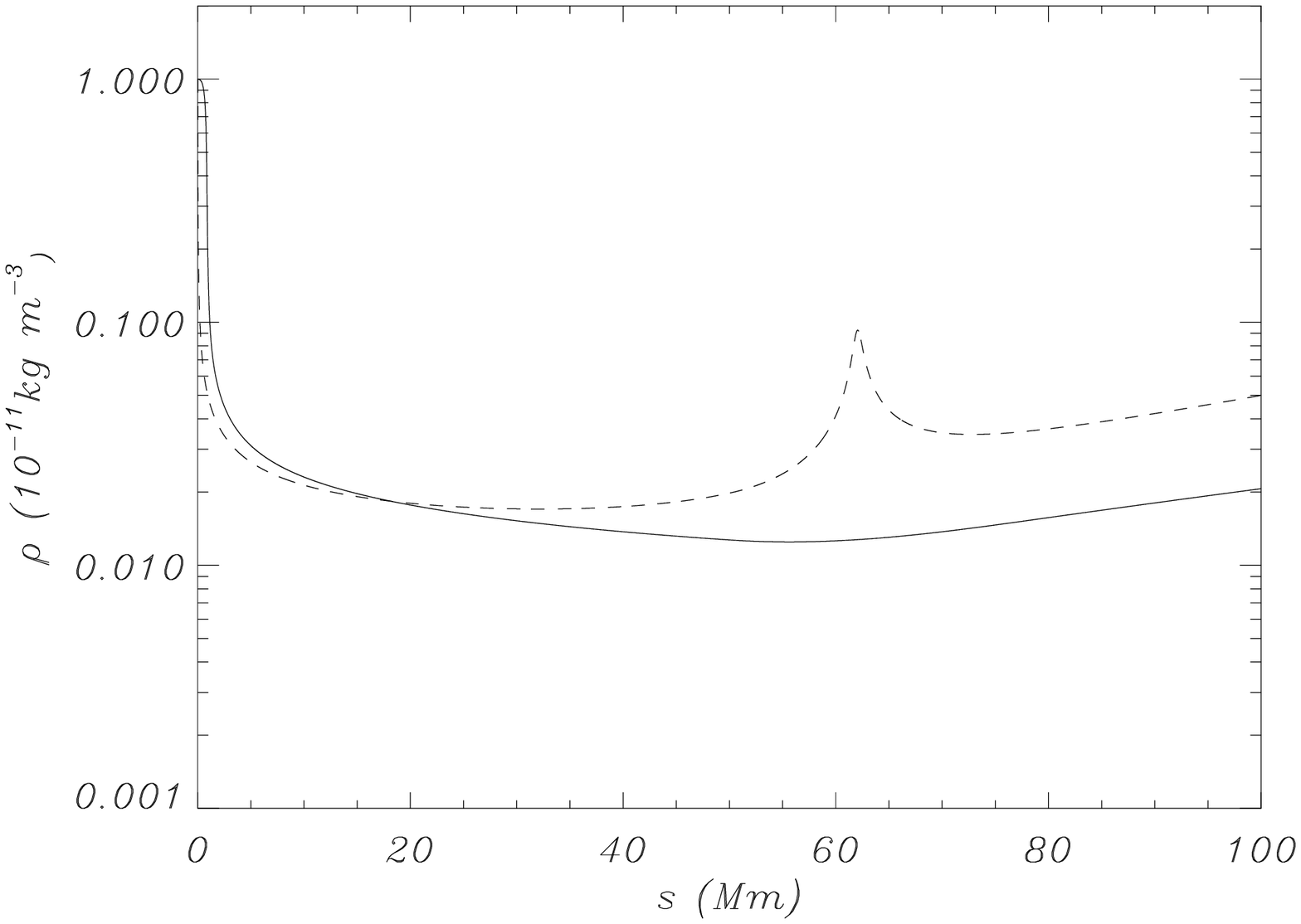}
\includegraphics[width=8cm]{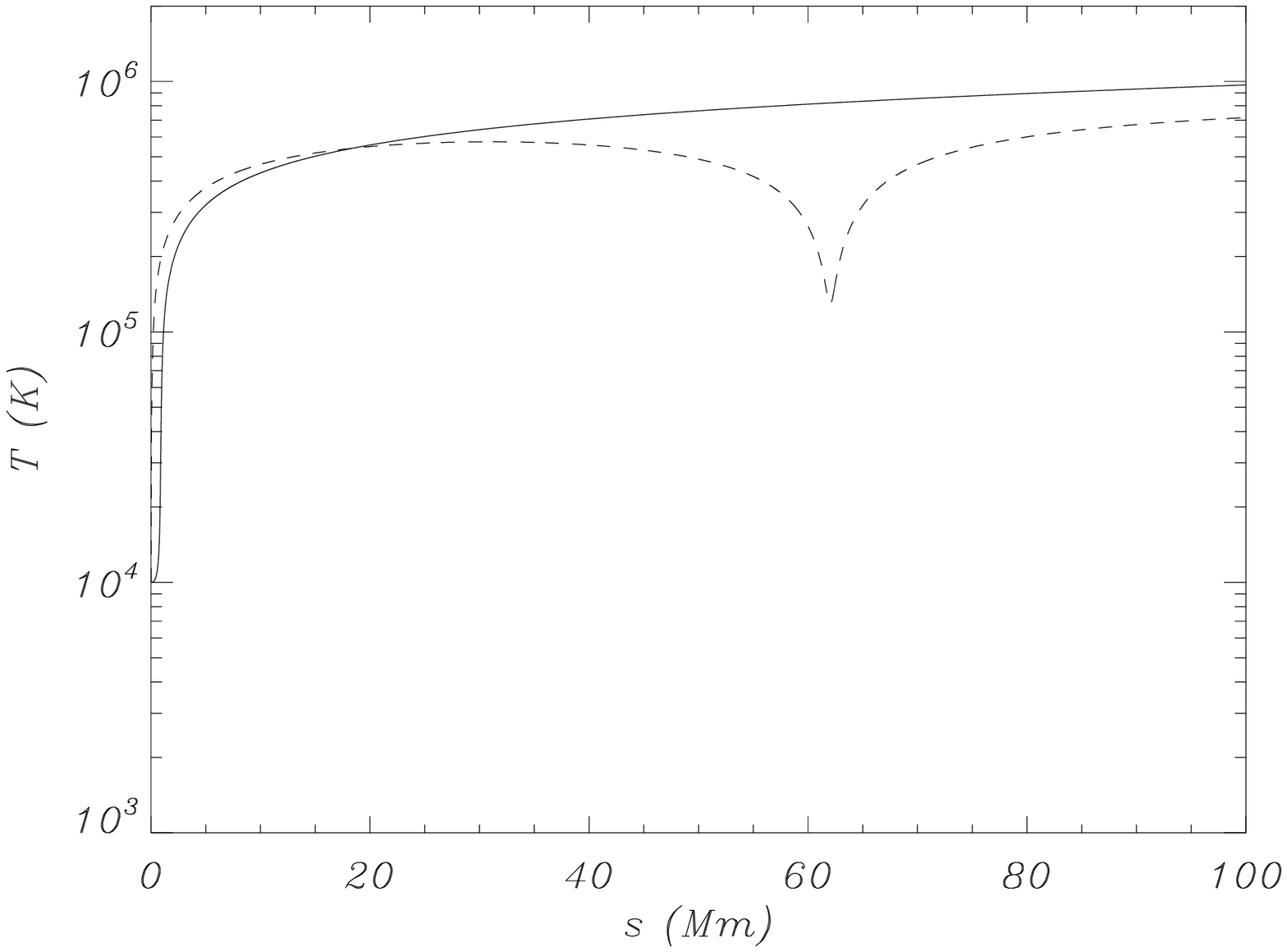} \caption{\small Hydrostatic and thermal
equilibrium along the field line with gravity. The continuous line corresponds
to Athays's radiation function, the dashed line represents the results for
Hildner's function, while the background heatings are $E_0=2.75\times 10^{-9}$ W
m$^{-3}$ and  $E_0=1.25\times 10^ {-6}$ W m$^{-3}$. In this
particular example $T_0=10^4\,\rm K$,  $\rho_0= 10^{-11}\,\rm kg
\, m^{-3}$ and the total length of the field line is $2 L=200\,\rm
Mm$. Model $A$ is used in this plot.}\label{comparg1} \end{figure}

\subsection{Non-constant gas pressure, gravity included}\label{sectgrav}

When gravity is introduced in the system gas pressure is no longer constant along
the magnetic field line. We concentrate hereafter in Model $A$. For large spatial scales we expect that
gravity can significantly alter the results in comparison to the constant pressure
case, and indeed this is the case. The numerical solution indicates that the second
thread solution (the one that was at $s\approx 45$ Mm in Fig.~\ref{comparg0}) 
for Hildner's function tends to have temperatures much below the
minimum at the thread centre ($s=0$) and the numerical integration of the equations fails to converge. The system does
not allow the periodicity found for the case with zero gravity for the same value of
the background heating. In this case no static solution is allowed in the system for
the selected parameters and a dynamical behaviour is expected because of the thermal
nonequilibrium.

However, by changing $E_0$ it is still possible to find, under some choice of parameters, a situation with two threads under mechanical and thermal
balance. An example is shown in Fig.~\ref{comparg1}. Thus, it is possible to have
some cold material in equilibrium balance but not located at the dips. This cold and
dense material does not reach the temperature and density values of the central thread
though.

In Fig.~\ref{comparg1}, density and temperature as a function of position along the
magnetic field under gravity is also represented for Athay's function. Interestingly,
in this case the  differences with respect to the zero gravity situation are not
large, essentially the density tends to increase near the footpoint as a result of the
presence of gravity since around the footpoint the gravity force is purely vertical
and  has the largest contribution (compare with the continuous line in the top panel
of Fig.~\ref{comparg0}). The thread length obtained in this case from the
simulations, hereafter denoted by $a$ (do not confuse with $l_{\rm th}$, the analytical approximation), is $1.7\,
\rm Mm$, and it is calculated using the position where the density derivative with $s$
has a maximum ($a$ is twice this value). We find an extended prominence corona
transition region that eventually matches a plasma that is close to coronal
conditions (densities of the order of $10^{-13}\,\rm kg\, m^{-3}$ and temperatures
around $10^6\,\rm K$). For these reasons we conclude that the obtained model is a good
representation of a thread. 

\begin{figure}[!hh] \center{\includegraphics[width=9cm]{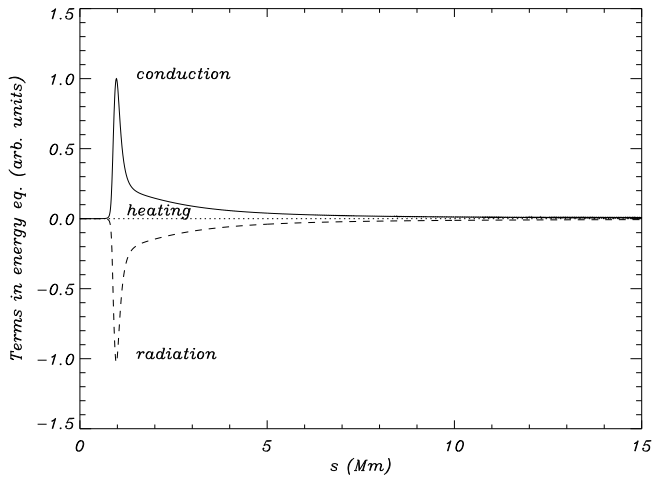}}
\caption{\small Terms in the energy equation, Eq.~(\ref{thermal}), as a function
of position at the thread body. The continuous line corresponds to the conduction term,
the dashed to the radiation losses while the dotted line represents the constant
background heating. The same parameters as in Fig.~\ref{comparg1} for Athay's
function have been used. Model $A$ is used in this plot.}\label{equilenerg} \end{figure}

In Fig.~\ref{equilenerg} the contribution of the different terms in the energy
equation is represented for the same parameters of Fig.~\ref{comparg1}. The
conduction term is always positive and is essentially balanced by the radiative
losses while the background heating is very small in this example. Both
conduction and radiation have a strong peak at the PCTR where density and
temperature change abruptly.

\begin{figure}[!hh] \center{\includegraphics[width=9cm]{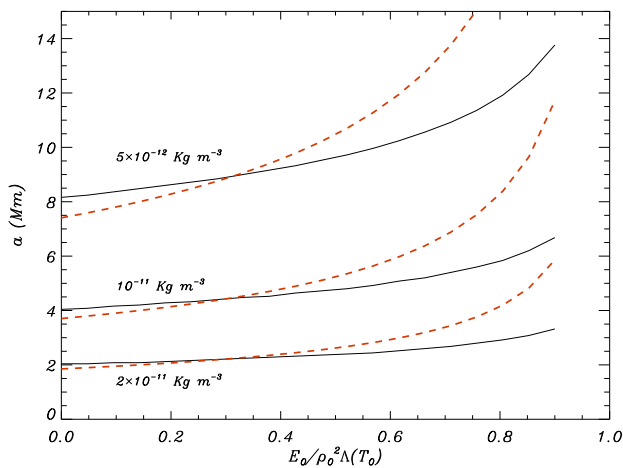}} 
\caption{\small Thread length, $a$, as a function of the heating constant, $E_0$, for
different reference densities $\rho_0$. In this plot the reference
temperature is $T_0=8\times10^3$ K and Athay's radiative loss has been
used. The  analytical approximation given by Eq.~(\ref{lthread}) is plotted with red dashed lines. Model $A$ is used in this plot.}\label{equilae0} \end{figure}

\subsection{A parametric survey}\label{paramsurv}

Once we know the main features of the solutions we study the influence of the
different parameters on the computed equilibrium for a wide range of values. For this
reason we have carried out a parametric study starting with the dependence on $E_0$.
Although from the results in Fig.~\ref{equilenerg} we may conclude that the role of
the background heating is small, it turns out that this parameter has a strong
influence on the obtained thread lengths, and as we will show in Paper II, this affects
significantly the damping times. This effect of $E_0$ on the equilibrium was already
noted by \citet{schmittdegen1995} and is further investigated here.

In
Fig.~\ref{equilae0} we have represented the numerically obtained thread lengths (twice the position where the density derivative with $s$ has a maximum) as a function of $E_0$ for
a specific choice of temperature ($8\times10^3$ K) at  the core of the thread. The
thread length decreases when the heating is reduced and the minimum thread length is
achieved for zero background heating. This agrees with the dependence of $l_{\rm th}$ on  $\rho_0^2\,\Lambda(T_0)-E_0$ in the denominator of Eq.~(\ref{lthread}).  Analytical approximations for the thread length are more difficult to obtain in
this case but note that the projection of gravity at $s=0$ is precisely zero,
therefore as a first approximation we can still use the definition of $l_{\rm th}$. The numerical results indicate that the longest thread lengths are obtained for
values of the heating tending to the radiative losses at the thread centre, as expected from Eq.~(\ref{lthread}) too. Nevertheless, the analytical approximation given by Eq.~(\ref{lthread}) and plotted in Fig.~\ref{equilae0} with red dashed lines overestimates the thread length in this last situation. It is worth to mention that there are numerical problems regarding convergence in the
calculation of the solutions for values of the heating constant near to $\rho_0^2\,
\Lambda(T_0)$, but the reason is clear from the denominator of $l_{\rm th}$.


We focus now on the dependence of the thread length on the length of the field lines. The three magnetic configurations, denoted by A, B and C, have been analysed and the total length of the
field lines has been changed in the range $80-200$ Mm. We have found that for Athay's radiative function the total length of
the threads, i.e. the parameter $a$, does not change much with the length of the magnetic field lines and the values are typically of the order of several Mm. Interestingly, the thread lengths are in the range of the reported in the
observations \citep[e.g.,][]{okamoto07,arreguietal2018}. The variation of the thread length with models A, B, and C is at most around 1\%. We conclude that the
geometry has some influence on the length of the threads but is not too significant. Note that this effect is not included in Eq.~(\ref{lthread}) since no gravity was assumed in the derivation of this equation (in our model the geometry of the field is only included through the projection of gravity along the magnetic field lines).


On the other hand, when Hildner's radiative function is considered, the thread lengths are short,
typically of the order of $150\, \rm km$ only. Again the magnetic field geometry has a weak
influence on the results in this case. The short lengths of the threads for Hildner's function were already reported by \citet{milne1979} and \citet{schmittdegen1995} and the cause is that the radiative losses are too high for typical thread temperatures. On the
contrary, Athay's radiative losses are significantly lower for typical thread
temperatures (see the comparison in Fig.~\ref{equilrad}) and this eventually leads to much longer threads.

\begin{figure}[!hh] \center{\includegraphics[width=9cm]{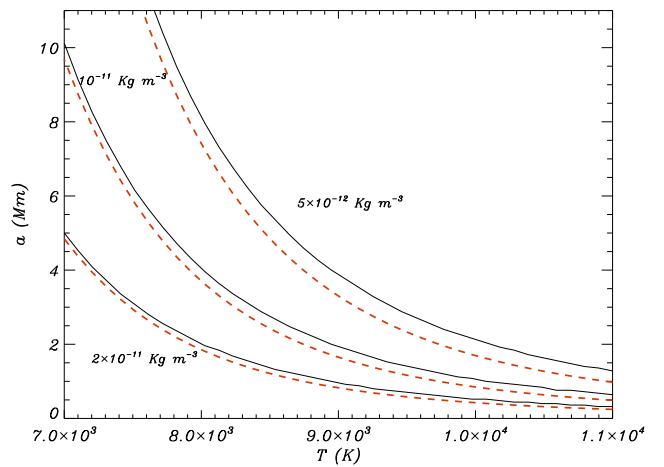}} 
\caption{\small Thread length, $a$, as a function of the central temperature for
different reference densities. The radiative function is based on
\citet{athay1986}. The  analytical approximation given by Eq.~(\ref{lthread}) is plotted with red dashed lines. Model $A$ is used in this plot.}\label{equilavt} \end{figure}

In Fig.~\ref{equilavt} the obtained thread length is
represented as a function of the central temperature for different reference
densities. In these calculations we have used Athay's function and have imposed
that $E_0=0$, therefore, as we have demonstrated in Fig.~\ref{equilae0} we
concentrate on the shortest threads. Figure~\ref{equilavt} indicates that cool
threads have longer lengths than  hot threads. Furthermore, light threads have larger sizes than heavy threads which is in  agreement again with Eq.~(\ref{lthread}) ($l_{\rm th}$ is inversely proportional to $\rho^2_0$), see the red dashed lines in Fig.~\ref{equilavt}.

\begin{figure}[!hh] \center{\includegraphics[width=9cm]{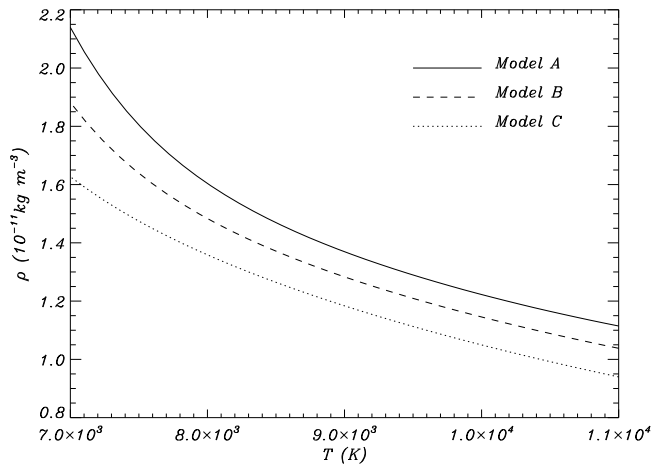}} 
\caption{\small Thread density at $s=0$, as a function of the central temperature for
models $A$, $B$, and $C$, matching a coronal temperature of 1 MK. A shooting method has been
used. The radiative function is the same as in \citet{athay1986}. }\label{equilshoot} \end{figure}

So far we have integrated the ordinary differential equations from $s=0$ to
$s=L$ because we have imposed three boundary conditions at $s=0$ (temperature,
density and zero derivative of temperature) and the problem is well defined. The
calculation of the maximum heating is straight forward since it involves the
density and temperature at the centre of the thread. Although in some cases we
have obtained coronal temperatures slightly below 1 MK in the corona this
approach provides reasonable results. However, it is also interesting to
investigate the situation where, apart from the thread temperature, the coronal
temperature is imposed and the integration provides the rest of the parameters. We
are mainly interested in the calculation of the density at the centre of the
thread that produces a perfect match for a coronal temperature at $s=L$. In this
case a shooting method is required to solve the differential equations. Now
the maximum heating constant is not known since it involves the thread density
that has to be determined from the solution of the equations. An iterative
approach is required to calculate the maximum $E_0$. To avoid this complication
we have concentrated on the situation $E_0=0$ and studied the dependence of the
results on temperature for the three different geometrical configurations. The
results are shown in Fig.~\ref{equilshoot}, where the obtained thread densities
are plotted as a function of the thread temperatures. The range of thread
temperatures for the three configurations agree well with the densities 
estimated from the observations \citep[e.g.,][]{tanberg1995,patvial2002}. The cooler the thread the higher the density is
the behaviour found in the configuration and it is what one would expect (using
the gas law) from a set of solutions that have essentially the same gas pressure
at the centre of the thread. We have investigated the situation when the heating
constant is different from zero, and again  it leads to longer threads than the
situation for $E_0=0$, as expected from Eq.~(\ref{lthread}) and demonstrated in Fig.~\ref{equilae0}.

\subsection{Matching a chromospheric layer} 

Up to now we have described a system composed of a cold and dense material
representing a thread connecting with a plasma under coronal conditions. A more
realistic model should include the connection with the chromosphere near the
footpoints of the magnetic field lines that are supposed to be anchored in the
photosphere. The physics of the chromosphere is complex and it is out
of the scope of this work to include a detailed modelling of this layer. However,
it is worth investigating the physical conditions required to have a layer akin
to the chromosphere in our simplified 1D system. We have already seen that it is
possible to obtain cold and dense plasma regions along the field lines depending
on the periodicity of the condensations that can have similar conditions to
chromospheric plasmas, especially under the presence of gravity. Nevertheless,
these chromospheric regions can occur anywhere along the field line, while we
are mostly interested in a chromosphere localised near the footpoints, i.e.,
near $s=\pm L$ in our model. The way to force the coronal part of the previously
obtained solutions to match with a chromosphere is to include a localised
heating near the footpoints. This approach has been used in the past by several
authors and the most
common form for the heating function used in the literature is,
\begin{eqnarray}\label{eodependence} E(s)=E_0\left(1+h_{\rm ch}\,
e^{(s-L)/\lambda}\right), \end{eqnarray}
where $\lambda$ is a spatial scale typically of the
order of 10 Mm and $h_{\rm ch}$ is a factor that is in the range 20-100 \citep[see for example,][]{dahlburgetal1998,karpenetal2001}. Using these
values in the spatially dependent heating function and solving the ordinary
differential equations we obtain a solution with a fast decrease in temperature
and a quick increase in density near the footpoint. However, the rapidly
changing nature of the chromospheric part of the solution requires, in general, a special treatment
of this layer \citep[see][for modifications in the conduction
coefficient to avoid such difficulties]{lionelloetal2009,johnstonbradshaw2019,zhouetal2021}. Nevertheless, the adaptive methods that are implicitly used in our numerical treatment do not fail to resolve the steep gradients in the solution. It is important to remark that the condition $E(s=0)<\rho_0^2\,\Lambda(T_0)$ still needs to be satisfied to have a thread-like solution. 

\begin{figure}[!hh] \centering \includegraphics[width=8.5cm]{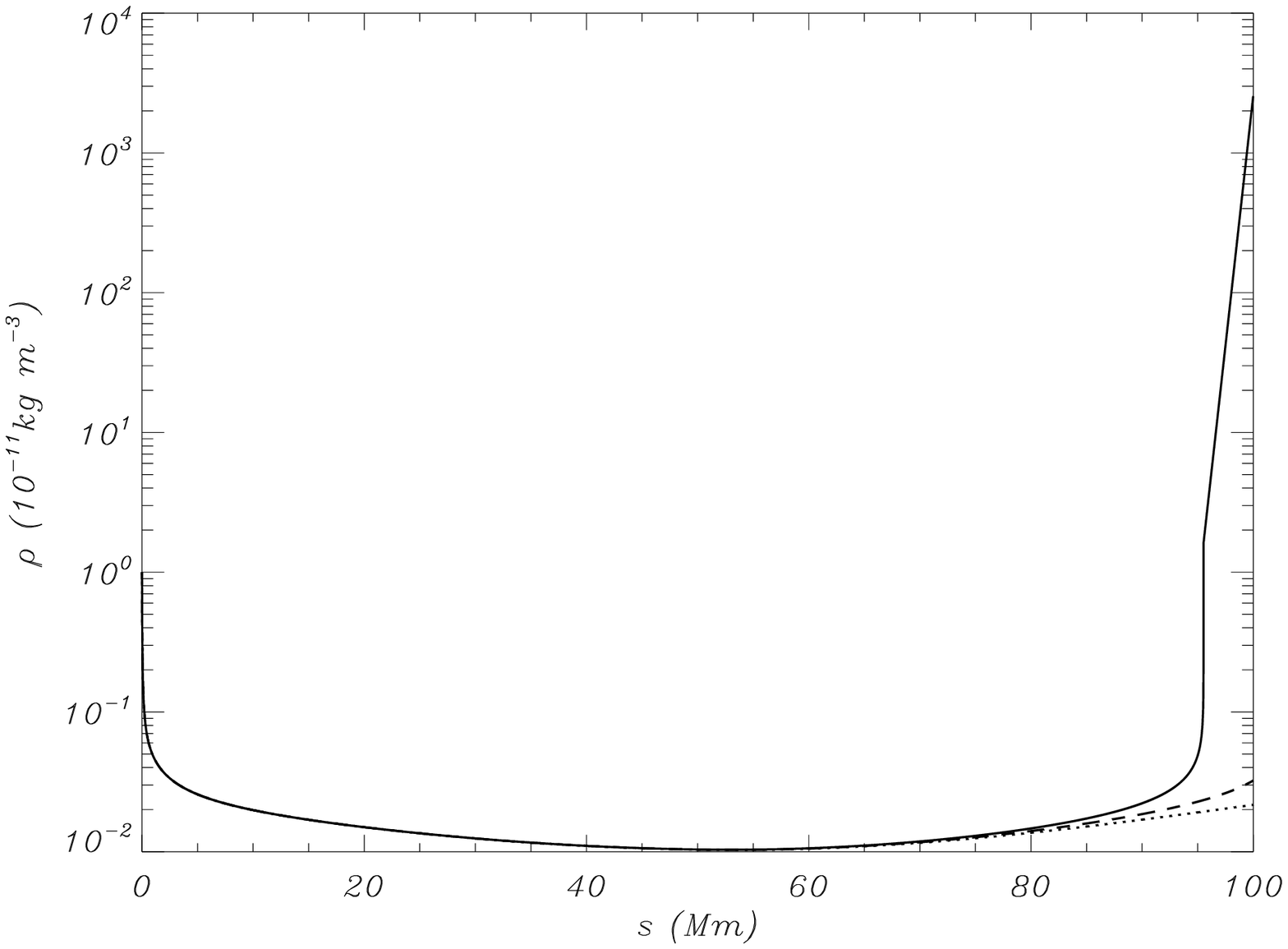}
\includegraphics[width=8.5cm]{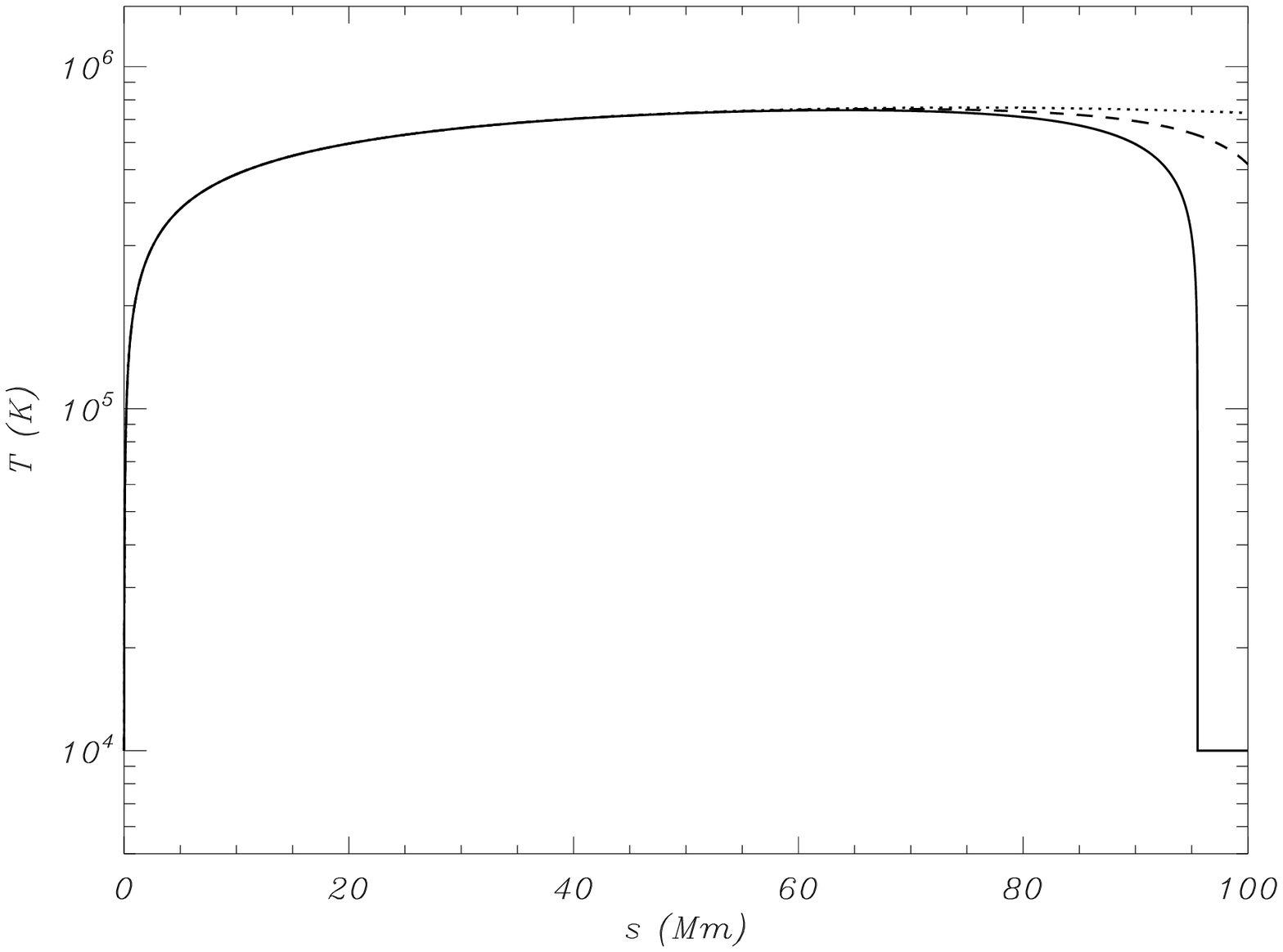}
\caption{\small Hydrostatic and thermal equilibrium along the field line with
gravity for Hildner's function. The background heating is $E_0=5\times
10^{-7}$ W m$^{-3}$ and  $h_{\rm ch}=35, 15, 0$ for the continuous, dashed, and dotted
lines. In
this particular example $T_0=10^4\,\rm K$,  $\rho_0=
10^{-11}\,\rm kg \, m^{-3}$ and the temperature threshold for the imposed
chromosphere is $10^4$ K. Model $A$ has been used in this example.}\label{chrom} \end{figure}

The approach used here to incorporate a chromospheric layer in the model
is to assume that at some point the temperature remains constant and takes typical chromospheric values \citep[see for example][]{mikicetal2013, karpenetal2001}. In this situation no
energy equation is required, gas pressure must be continuous at this point and
the presence of gravity leads to an exponentially growing density and pressure as
we move downwards in the chromospheric part. The integration of the equations is performed from the thread centre to the chromosphere. An example of such computed equilibrium is
shown in  Fig.~\ref{chrom} for the case $h_{\rm ch}=35$. In this equilibrium a temperature  threshold  of
$10^4$ K has been imposed in the isothermal chromospheric layer which starts at
$s=95$ Mm and its thickness is therefore 5 Mm (larger than the real thickness which is typically around 2 Mm). The exponential increase in density (straight line in the logarithmic
scale of the plot) is clear in the top panel of the figure and leads to mean chromospheric density values.  

Using Hildner's function and for comparison purposes we have included in Fig.~\ref{chrom} two solutions with
reduced localised heating to visualise the nature of the solutions. For the situations
with $h_{\rm ch}=15$ and  $h_{\rm ch}=0$ we find how temperature decreases near the footpoint but is still
above the temperature threshold ($10^4\,\rm K$) to have a chromosphere. An interesting conclusion from Fig.~\ref{chrom} is that the presence of footpoint heating does not affect the thread properties and the solution is essentially the same around $s=0$ for the three cases. This is also true when Athay's radiative function is considered. It is worth mentioning that multiple condensations, discussed in Sect.~\ref{sectgrav},  are not present in Fig.~\ref{chrom} because the heating constant is always lower than the one used in  Fig.~\ref{comparg1}.

The model developed in this section including a 
dense and cold chromosphere near the footpoint will be useful for understanding
the effect of this layer on the attenuation of the waves (Paper II) produced by the
possible mechanism of wave reflection at the chromosphere and wave
leakage if the boundary is open.

\section{Partially ionised plasma}\label{partialion}

Let us assume that the plasma is not fully ionised. Based on 1D non-LTE
radiative transfer models, \citet{heinzeletal2014} calculated among other
parameters, the ionisation degree in several prominence slabs. In particular, \citet{heinzeletal2015} provide tables for the
ionisation degree for different temperatures and pressures at the prominence.
The idea here is to use these values in our calculations of the equilibrium and
study how our models are modified by the presence of neutrals. We are again in
the situation where the description using an optically thin plasma is not fully correct but it can be considered as a starting point.

The plasma is assumed to be composed of hydrogen and helium. The abundance of helium is $10\%$ and is not ionised. The ionisation degree is defined
here as $i=n_{\rm e}/n_{\rm H}$ where $n_{\rm e}$ is the electron density and $n_{\rm
H}$ the total hydrogen density ($n_{\rm H_{I}}+n_{\rm p}$). The total particle
number is defined as $N=n_{\rm H}+n_{\rm He}+n_{\rm e}$ and using the previous
definitions and the helium abundance ($n_{\rm He}=0.1\, n_{\rm H})$  it is written as 
\begin{eqnarray}\label{Ntot}
N=n_{\rm e}\left(1+\frac{1.1}{i}\right).
\end{eqnarray}
Using the ideal gas law we find that the electron density in terms of pressure,
temperature and ionisation degree is (now we write the explicit dependence of the variables)
\begin{eqnarray}\label{ne}
n_{\rm e}(s)=\frac{p(s)}{k_{\rm B} T(s)} \left(\frac{i(p(s),T(s))}{i(p(s),T(s))+1.1}\right).
\end{eqnarray}
The ionisation degree, $i$, depends on $p$ and $T$ and it is calculated from
Table 1 of \citet{heinzeletal2015}. To simplify things hereafter we write $i(p(s),T(s))$ as $i(s)$.

For the total density we have that $\rho=n_{\rm H} \,m_{\rm p}+n_{\rm He}\, 4\, m_{\rm
p}+n_{\rm e}\, m_{\rm e}$ (being $m_{\rm p}$ and $m_{\rm e}$ the proton and electron
masses). It reduces to the following expression when the
electron mass is neglected in front of the proton mass,
\begin{eqnarray}\label{rhopartialion}
\rho(s)=\frac{m_{\rm p}}{k_{\rm B}}\frac{p(s)}{T(s)} \left(\frac{1.4}{i(s)+1.1}\right).
\end{eqnarray}
Now the term in the brackets plays the role of $\tilde{\mu}$ in Eq.~(\ref{gaslaw}).
In the partially ionised situation it is more convenient to solve the equilibrium
equations for gas pressure and temperature because the ionisation degree
calculated in \citet{heinzeletal2015} depends
on these two magnitudes.

The equation for hydrostatic equilibrium, Eq.~(\ref{hydrop}), is written in
terms of pressure and temperature because density has been eliminated using
Eq.~(\ref{rhopartialion}). Partial ionisation changes the equation for
thermal equilibrium in the conduction and radiation terms. The factor in front of the radiative
term is now 
\begin{eqnarray}\label{rad}
n_{\rm e}(s)\, n_{\rm H}(s)=\frac{n_{\rm e}^2(s)}{i(s)}.
\end{eqnarray}
In the conductivities the contribution of neutrals has to be added to the
electron contribution. We have that according to \citet{soleretal2010b,soleretal2012},  
\begin{eqnarray}\label{conde}
\kappa_{\rm e}(s)&=& \kappa_0\, \xi_{\rm p}(s)\, T^{5/2}(s),\\
\kappa_{\rm n}(s)&=&\left( \kappa_1 \,\xi_{\rm
H_{I}}(s)+\kappa_2\, \xi_{\rm He_{I}}\right) T^{1/2}(s),
\end{eqnarray}
where the relative density of species are in our case, $\xi_{\rm p}(s)=i(s)/1.4$,
$\xi_{\rm H_{I}}(s)=(1-i(s))/1.4$ and $\xi_{\rm He_{I}}=0.4/1.4$. For the
conductivities we have that $\kappa_0=1.1\times 10^{-11}$ W m$^{-1}$ K$^{-7/2}$, $\kappa_1=2.24\times 10^{-2}$ W m$^{-1}$ K$^{-3/2}$, and $\kappa_2=3.18\times 10^{-2} $ W m$^{-1}$ K$^{-3/2}$. The effective
conduction coefficient is now the sum of electron and neutral conductivities
\begin{eqnarray}\label{condtot} 
\kappa_\parallel(s)=\kappa_{\rm e}(s)+\kappa_{\rm n}(s).
\end{eqnarray}
The conduction term involves spatial derivatives of this coefficient. The coefficient $\kappa_\parallel$ depends on temperature and ionisation degree, and this last parameter
also depends on temperature and pressure. Regarding the ionisation, from the practical point of view we have adjusted a second order polynomial to
the function $i(p,T)$ given in  \citet{heinzeletal2015} (table for an altitude of 20 Mm). The calculation of the partial derivatives of $i$ with
$p$ and $T$ are straight forward using the polynomial fit.

\begin{figure}[!hh] \center{\includegraphics[width=9cm]{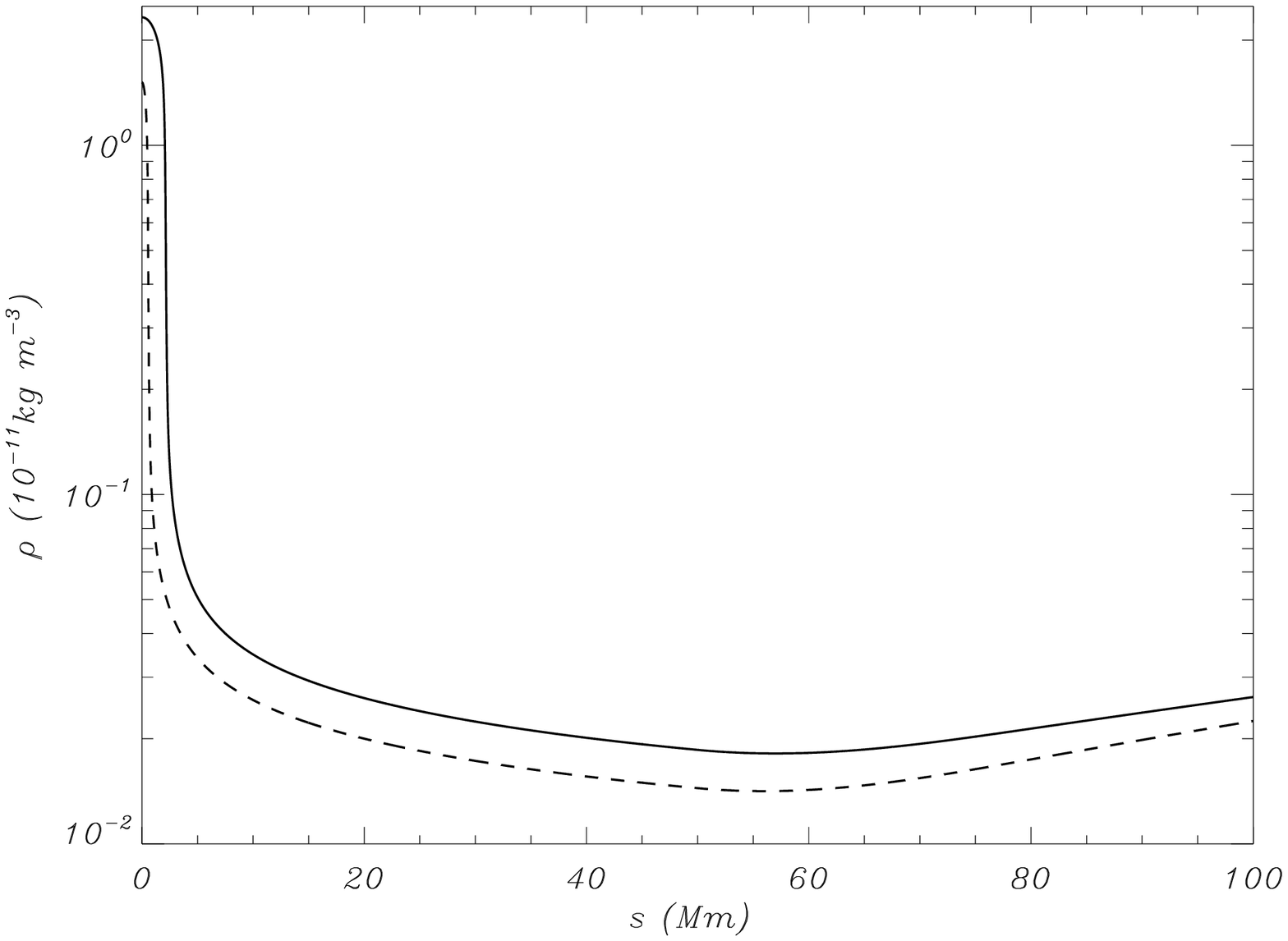}} 
\center{\includegraphics[width=9cm]{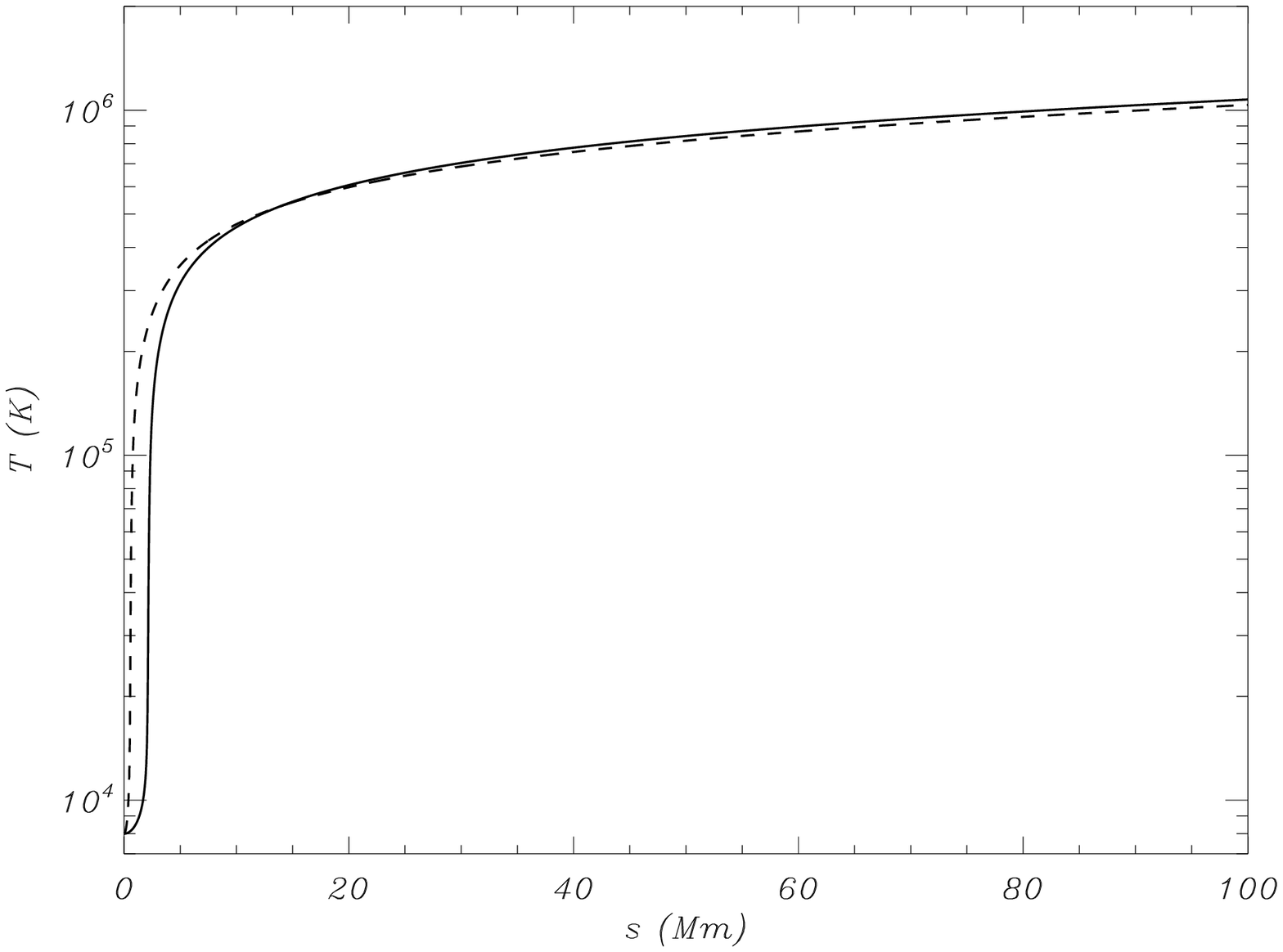}} 
\center{\includegraphics[width=9cm]{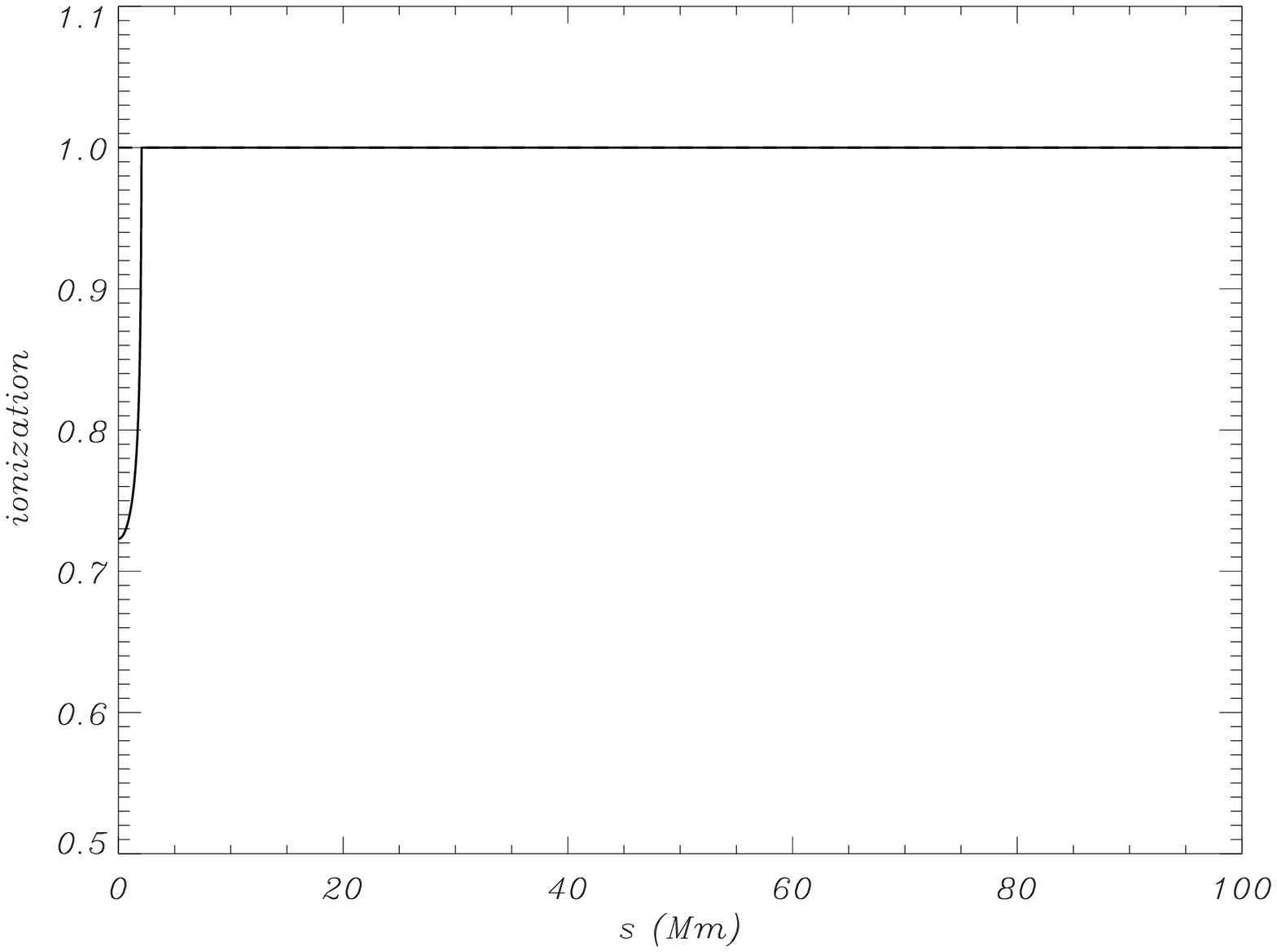}}  \caption{\small Hydrostatic and
thermal equilibrium along a thread under partial ionisation (continuous lines) and
full ionisation (dashed lines). In this particular example, temperature
($T_0=8\times10^3$ K) and pressure at ($s=0$) are the same for the partially and fully
ionised situation. In these solution we have imposed that $E_0=0$. The radiative
function is the same as in \citet{athay1986}. Model $A$ has been used in this plot.}\label{equilioniz} \end{figure}

We have repeated some of the previous calculations including the effect of partial
ionisation through the change in the radiation and conduction terms and the
modification of the ideal gas law. Since we have shown that the inclusion of a chromosphere has little effect on the conditions within the threads we do not include this pseudo-layer in the present calculations.

The main result is that partial ionisation
significantly increases the size of the thread. An example is shown in
Fig.~\ref{equilioniz},  where the length of the thread is four times longer than in the
fully ionised case. We have to bear in mind that now the initial values of the thread
parameters at $s=0$, required to perform the integration of the differential equations,
are different to those in the fully ionised case because of the change introduced in
the gas law by partial ionisation (see Eq.~(\ref{rhopartialion})). In the present
example temperature and gas pressure are the same at the thread centre but the
reference density is different. In Fig.~\ref{equilioniz} (bottom panel) we can see
how the ionisation degree changes inside the thread because of the implemented model
of \citet{heinzeletal2015}. The ionisation degree raises smoothly from 0.72 at the
thread centre to  1 at the edge of the thread where high (coronal) temperatures are
achieved.  

The increment in the thread length produced by partial ionisation is a
consequence of the changes in thermal conduction, which is larger owing to the
contribution of neutrals on the conductivities. For a better understanding of this feature we
proceed as in the fully ionised situation when gravity is absent. Using the same approximation as in Eq.~(\ref{tempexp}) we find that under the
presence of partial ionisation the characteristic spatial scale of the thread is
\begin{eqnarray}\label{lthreadio}
l_{\rm th}=\sqrt{2\frac{
\kappa_0 \frac{i_0}{1.4} T^{7/2}_0 +\left(\kappa_1\frac{1-i_0}{1.4}
+\kappa_2\frac{0.4}{1.4}\right) T^{3/2}_0 } {\rho_0^2
\Lambda(T_0)\frac{i_0}{1.4^2}-E_0}},
\end{eqnarray}
where $i_0=i(p_0,T_0)$ is the ionisation degree at the centre of the thread. The
value of $i_0$ is typically around 0.72 but the reference density, $\rho_0$, also depends on $i_0$ (see Eq.~(\ref{rhopartialion})). Indeed, it is not difficult to calculate the ratio of the radiation term for partial ionisation relative to full ionisation for constant reference pressure, $p_0$ and temperature, $T_0$. For partial ionisation the reference density is 
\begin{eqnarray}\label{rhopartialion0}
\rho_{\rm 0p}=\frac{m_{\rm p}}{k_{\rm B}}\frac{p_0}{T_0} \left(\frac{1.4}{i_0+1.1}\right),
\end{eqnarray}
while for full ionisation
\begin{eqnarray}\label{rhopartialion0f}
\rho_{\rm 0f}=\frac{m_{\rm p}}{k_{\rm B}}\frac{p_0}{T_0} \left(\frac{1}{2}\right).
\end{eqnarray}
The ratio of the two corresponding radiation terms is
\begin{eqnarray}
\frac{\rho_{\rm 0p}^2\,
\Lambda(T_0)\frac{i_0}{1.4^2}}{\rho_{\rm 0f}^2\,
\Lambda(T_0)}=\frac{4\, i_0}{\left(i_0+1.1\right)^2}.
\end{eqnarray}
This ratio is always smaller than one, meaning that radiative losses under partial ionisation are reduced in comparison with the fully ionised case. Since this term appears in the denominator of Eq.~(\ref{lthreadio}) it means that it produces an increase in the thread length.

\begin{figure}[!hh] \center{\includegraphics[width=9cm]{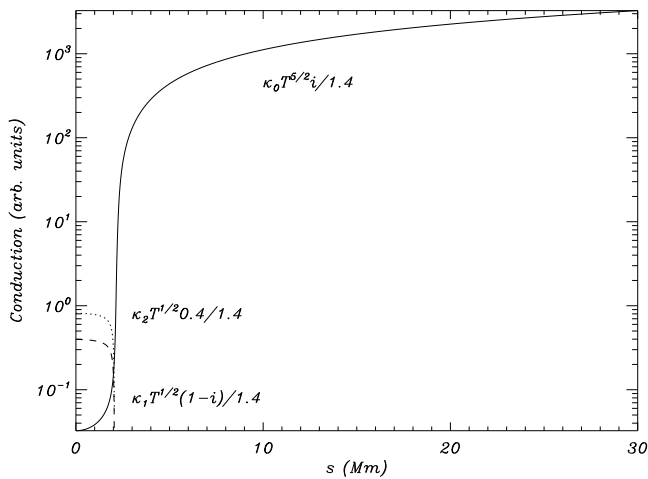}} \caption{\small Conduction terms as a function of position (in the range 0 to 30
Mm for visualisation purposes). Conduction by neutral H (dashed line) goes to
zero at the edge of the thread where the plasma is fully ionised. Conduction by
neutral He (dotted line) dominates inside the thread and becomes zero in the 
corona where there is conduction by electrons only (continuous
line).  Same parameters as in Fig.~\ref{equilioniz} have been used.}\label{conducts} \end{figure}

\begin{figure}[!hh] \center{\includegraphics[width=9cm]{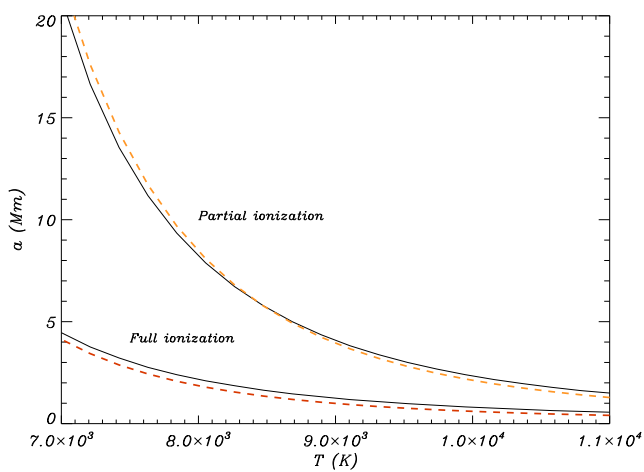}} 
\caption{\small Thread length as a function of temperature under partial ionisation and full ionisation. In these solution we have imposed
that $E_0=0$. The radiative function is the same as in \citet{athay1986}. The  analytical approximations given by Eqs.~(\ref{lthreadio}) (partial ionisation)  and (\ref{lthread}) (full ionisation) are plotted with orange and red dashed lines. Same parameters as in Fig.~\ref{equilioniz} have been used.}\label{widthpartial}
\end{figure}

For thermal conduction we need to evaluate the three terms that appear in the
numerator of Eq.~(\ref{lthreadio}). The sum of the three terms is typically
around 4.5 times the value for the  fully ionised situation. This increment leads to
longer threads under partial ionisation. It turns out that the increased conduction 
under partial ionisation is mostly produced by the presence of  neutral helium. This is
visualised in Fig.~\ref{conducts} where the different conduction terms are plotted
as a function of position for a typical case. Inside the thread, conduction by neutrals
dominates conduction by electrons. The contribution of neutral helium is the largest,
although its abundance is just $10\%$. Conduction by neutral H goes to zero as we
approach the edge of the thread where plasma is fully ionised. In the coronal medium
the contribution of neutral He is artificially forced to go to zero since at coronal temperatures is fully ionised. Neutral He is therefore responsible, together with the reduced
radiation, for the rise of the obtained thread lengths.


Finally, Fig.~\ref{widthpartial} shows how the thread length changes with the
central temperature under partial ionisation. The behaviour is the
same as in Fig.~\ref{equilavt} for the fully ionised case, and for comparison
purposes we have also represented the results for full ionisation for
the same reference pressure at $s=0$. Interestingly, as the central temperature increases the two curves approach each other since the larger the temperature the larger the ionisation degree in the prominence. We find again longer threads under the presence
of partial ionisation, being the difference relative to full ionisation a factor
that varies from 5.3 to 2.6 in the range of temperatures of the plot. The analytical approximation given by Eq.~(\ref{lthreadio}) is also plotted in the figure and the agreement with the numerical result is remarkable.

\section{Summary and conclusions}

In the present work we have analysed the features of one-dimensional equilibrium thread models under hydrostatic and thermal balance. We have started with the situation without gravity and have progressively increased the complexity of the model. This has allowed us a better comprehension of the results. Gravitational stratification, the presence of a chromosphere and finally partial ionisation effects have been incorporated to our model. The main outcome of our study is summarised in the following:

   \begin{enumerate} 
   
\item The value of the background heating in comparison with the radiative
losses at the centre of the thread is crucial to lead to thread-like solutions
surrounded by coronal plasma. Only if $E_0<\rho_0^2\,\Lambda(T_0)$  static, cold
and dense plasma threads under thermal equilibrium exist. In the opposite situation when the background heating is higher than the radiative losses at the centre of the structure only loop-like solutions are achieved.

\item Several thread-like condensations are in principle possible along the field lines and
not necessarily located at the dips of the magnetic field. We have found that the physical origin of the secondary condensations
is related to the critical length introduced by \citet{field1965} and related to the characteristic spatial scale for thermal stability.

\item The presence of gravity in the model can produce that the secondary
condensations collapse and no equilibrium configuration on a given length of the
magnetic field line is achieved. Again the value of the background heating plays
a major role in the behaviour of secondary thread-like solutions, located in general near the footpoints. The gravitational stability of the obtained solutions needs to be investigated and it will be addressed in Paper II.

      \item A parametric survey has been carried out to understand the dependence
      of the thread length, on the different values of the parameters. The
      geometry of the field lines is not especially important but the radiative
      losses for low temperatures are crucial to obtain realistic thread
      lengths. Athay's radiative function, with reduced losses under typical
      thread conditions (see Fig.~\ref{equilrad}), is the most suitable choice from the ensemble of radiative losses
      analysed in the present work. In any case since the plasma is optically thick under typical prominence/thread conditions the incorporation of more realistic physics requires to properly solve the radiative transport problem.

     \item We have derived a simple analytical expression for the characteristic spatial scale of the thread, $l_{\rm th}$, under static equilibrium that explains the dependence of the computed thread lengths, $a$, with the different parameters of the model, including partial ionisation. This could be used in a novel way, for example, to infer the ratio between radiation and heating if the length of the thread and the central temperature and density are known from observations.

    \item Significantly long threads are obtained when partial ionisation is
   present. This is a consequence of the reduced radiation and increased conduction produced by
   neutral helium in comparison to the fully ionised case. It is assumed that helium is
   totally neutral in the thread, but this is not completely true under real
   conditions \citep[see][]{soleretal2010b} and in reality this may decrease the
   length of the threads. This needs to be addressed in the future.

      \item The connection of the thread-corona system with a chromosphere is
      obtained when heating around a certain threshold is localised near the
      footpoints, in agreement with \citet{dahlburgetal1998}. The chromospheric model used here (a  stratified and isothermal
      layer under full ionisation) is a first approximation and more physics needs to be included in future
      studies. However, the presence of a chromosphere is interesting for the applications regarding the damped oscillations that will be discussed in Paper II.

    \item Related to the previous point, we have shown that localised footpoint heating does not significantly alter the temperature and density of the cold plasma representing the thread relative to the case without footpoint heating, at least for the parameters considered in this work. In our model the localised heating essentially leads to the  existence of a chromospheric layer only.
   
   \end{enumerate}

The numerical solutions presented in this work will be used to compute the corresponding eigenmodes to compare the damping rates of our calculations and the reported in the observations (Paper II). First, this will allow us to better understand the damping mechanism of the oscillations and second this information will be used as a method to 
constrain or to infer some modifications to the  models \citep[see][]{anzerheizel2008A} and most likely about the dependence of the radiative losses on temperature. The comparison of theory and observations will be used for seismological purposes.
   
\begin{acknowledgements} We acknowledge the support from grant AYA2017-85465-P
(MINECO/AEI/FEDER, UE), to the Conselleria d'Innovaci\'o, Recerca i Turisme del
Govern Balear, and also to IAC$^3$. We are grateful to the ISSI Team led by Manuel
Luna ``Large-Amplitude Oscillations as a Probe of Quiescent and Erupting Solar
Prominences'' for inspiring this work in the fruitful meetings held in Bern in 2018
and 2019. We also thank ISSI-Beijing for hosting the team "The eruption of solar filaments and the associated mass and energy transport” led by J. C. Vial and P. F. Chen where part of the results of this paper were presented. We are grateful to Rafel Prohens from the Departament de Ci\`encies Matem\`atiques i Inform\`atica, Universitat de les Illes Balears (UIB) for his advise on phase analysis of nonlinear differential
equations. We thank the anonymous referee for their useful comments that helped to improve the manuscript.\end{acknowledgements}

\bibliographystyle{aa}      
\bibliography{jaume}   

\end{document}